\documentclass[draft]{agujournal2018}
\usepackage{apacite}
\usepackage{url} 
\usepackage[version=4]{mhchem}
\usepackage[pdftex]{hyperref}
\usepackage{rotfloat}
\usepackage{amsmath}
\usepackage{amssymb}
\usepackage{graphicx}
\usepackage{lscape}
\usepackage{float}

\draftfalse

\journalname{Geochemistry, Geophysics, Geosystems}

\begin{document}

%
%


\title{Nitrogen Oxide Concentrations in Natural Waters on Early Earth}

%
%




\authors{Sukrit Ranjan\affil{1,*}, Zoe R. Todd\affil{2},  Paul B. Rimmer\affil{3,4},  Dimitar D. Sasselov\affil{2}, Andrew R. Babbin \affil{1}}

\affiliation{1}{Massachusetts Institute of Technology, Department of Earth, Atmospheric, and Planetary Sciences}
\affiliation{2}{Harvard University, Department of Astronomy}
\affiliation{3}{MRC Laboratory of Molecular Biology}
\affiliation{4}{University of Cambridge, Cavendish Astrophysics}

\affiliation{1}{77 Massachusetts Avenue, Cambridge, MA 02139}
\affiliation{2}{60 Garden Street, Cambridge, MA 02138}
\affiliation{3}{Francis Crick Ave, Cambridge CB2 OQH, UK}
\affiliation{4}{JJ Thomson Avenue, Cambridge CB3 0HE, UK}





\correspondingauthor{Sukrit Ranjan}{Email: sukrit@mit.edu. T: 617-253-6283. F: 617-324-2055. Mailing: 77 Massachusetts Avenue, Room 54-1719, Cambridge, MA 02139}




\begin{keypoints}
\item Nitrate and nitrite (NO$_X^-$) are relevant to prebiotic chemistry. Past work has argued these molecules were abundant in the early ocean.
\item Fe$^{2+}$ and UV suppress [NO$_X^-$] to much lower concentrations than previously though in the ocean. [NO$_X^-$] could have been higher in ponds.
\item Most NO$_X^-$ should have been nitrate. Prebiotic chemistries that use nitrate are more plausible than those that use nitrite. 
\end{keypoints}

%
%


\begin{abstract}
A key challenge in origins-of-life studies is estimating the abundances of species relevant to the chemical pathways proposed to have contributed to the emergence of life on early Earth. Dissolved nitrogen oxide anions (\ce{NO_{X}^{-}}), in particular nitrate (\ce{NO_3^-}) and nitrite (\ce{NO_2^-}), have been invoked in diverse origins-of-life chemistry, from the oligomerization of RNA to the emergence of protometabolism. Recent work has calculated the supply of \ce{NO_{X}^{-}}~from the prebiotic atmosphere to the ocean, and reported steady-state [\ce{NO_{X}^{-}}] to be high across all plausible parameter space. These findings rest on the assumption that \ce{NO_{X}^{-}}~is stable in natural waters unless processed at a hydrothermal vent. Here, we show that \ce{NO_{X}^{-}}~is unstable in the reducing environment of early Earth. Sinks due to UV photolysis and reactions with reduced iron (\ce{Fe^{2+}}) suppress [\ce{NO_{X}^{-}}] by several orders of magnitude relative to past predictions. For pH$=6.5-8$ and $T=0-50^\circ$C, we find that it is most probable that [\ce{NO_{X}^{-}}]$<1~\mu$M in the prebiotic ocean. On the other hand, prebiotic ponds with favorable drainage characteristics may have sustained [\ce{NO_{X}^{-}}]$\geq 1~\mu$M. As on modern Earth, most \ce{NO_{X}^{-}}~on prebiotic Earth should have been present as \ce{NO_3^-}, due to its much greater stability. These findings inform the kind of prebiotic chemistries that would have been possible on early Earth. We discuss the implications for proposed prebiotic chemistries, and highlight the need for further studies of \ce{NO_{X}^{-}}~kinetics to reduce the considerable uncertainties in predicting [\ce{NO_{X}^{-}}] on early Earth.
\end{abstract}

%
%
\section{Introduction\label{sec:intro}}
A key challenge for origin-of-life studies is determining the range of environmental conditions  on early Earth under which life arose. Knowledge of these environmental conditions informs development of theories of the origin of life, and enables assessment of the plausibility and probability of postulated prebiotic chemistries (e.g., \citet{Urey1952, Corliss1981, Bada1994, Ruiz-Mirazo2014, Todd2018, Xu2018}). Consequently, extensive work has been done to place constraints on the prebiotic environment, including but not limited to the availability of liquid water, the redox state of the atmosphere, the UV irradiation environment, the pH and temperature of the early oceans, the physico-chemical conditions at deep-sea hydrothermal vents, and the availability of sulfidic anions \citep{Farquhar2000,Mojzsis2001,Martin2008, Ranjan2017a,Halevy2017, Krissansen-Totton2018ocean,Ranjan2018}. 

An important prebiotic environmental factor is the abundance of fixed nitrogen species in natural waters on early Earth. Dinitrogen's high-energy triple bond renders it highly nonreactive, meaning that nitrogen generally must be ''fixed" into its reduced or oxidized forms (e.g., \ce{NO_2^-}, \ce{NO_3^-}, NH$_4^+$) to be useful to biology or prebiotic chemistry. Consequently, it is unsurprising that nitrogen fixation is thought to be ancient \citep{Fani2000, Zehr2002, Canfield2010, Stueken2015, Zerkle2017}. Prebiotic chemists have been especially interested in nitrate (\ce{NO_3^-}) and nitrite (\ce{NO_2^-}), the oxidized anions of nitrogen (\ce{NO_{X}^{-}}). These molecules are high-potential electron acceptors, have played a key role in  microbial metabolisms since at least the Archaean \citep{Canfield2010}, and have been hypothesized to have been involved in the first metabolic pathways, e.g, the oxidation of methane and hydrogenation of CO$_2$ at deep-sea vents \citep{Ducluzeau2009, Nitschke2013, Shibuya2016}. These molecules have also been invoked for the non-enzymatic synthesis and replication of oligonucleotides, in surficial (lake/pond) settings \citep{Mariani2018}. Crucially, \ce{NO_{X}^{-}}~can be abiotically synthesized by the thermal decay of molecules like HNO, which are produced by high-energy events like lightning in an N$_2$-CO$_2$ atmosphere \citep{Kasting1981, Mancinelli1988, Navarro-gonzalez1998, Ardaseva2017}, meaning these molecules may have been available for prebiotic chemistry on early Earth. 

Motivated by the potential prebiotic relevance of \ce{NO_3^-}~and \ce{NO_2^-}, a number of studies have aimed to constrain their concentrations on early Earth. \citet{Mancinelli1988} pointed out that atmospherically-generated \ce{NO_{X}^{-}}~would form \ce{NO_2^-}~and \ce{NO_3^-}~in the prebiotic ocean and accumulate, and that solubility concerns would not limit the accumulation. However, \citet{Mancinelli1988} did not quantify the concentrations to which \ce{NO_{X}^{-}}~could accumulate. \citet{Wong2017} conducted atmospheric modeling, combining 3D General Circulation Model (GCM) estimates with 1D photochemical models to estimate the supply of \ce{NO_{X}^{-}}~to the oceans due to lightning. They identified the key variable controlling the \ce{NO_{X}^{-}}~supply to be the CO$_2$ partial pressure (pCO$_2$). Under the assumption that the sole sink on \ce{NO_{X}^{-}}~was destruction at high-temperature hydrothermal vents, \citet{Wong2017} computed [\ce{NO_{X}^{-}}] in the bulk ocean to be $\geq 10 \mu$M, and [\ce{NO_{X}^{-}}]$=20$ mM for pCO$_2=1$ bar. \citet{Laneuville2018} conducted a systems model of the prebiotic nitrogen cycle, including cometary delivery, impact synthesis, and lightning as sources of fixed nitrogen, and destruction at hydrothermal vents as the sole sink of oceanic \ce{NO_{X}^{-}}. They calculated [\ce{NO_{X}^{-}}]$\approx1\mu\text{M} - 10\text{mM}$ in the bulk prebiotic oceans, depending on a number of variables including atmospheric nitrogen fixation rate. These large ranges are due to the wide range of possible \ce{NO_{X}^{-}}~supply. Both \citet{Wong2017} and \citet{Laneuville2018} suggest that the prebiotic oceans should have had $\geq 1 \mu$M [\ce{NO_{X}^{-}}], thought to be adequate for prebiotic chemistry. For comparison, on modern earth, bioavailable fixed \ce{NO_{X}^{-}} achieves maximum concentrations of $\sim40~\mu$M in the deep Pacific \citep{Zehr2002, Olsen2016}.  \ce{NO_{X}^{-}}  in the modern ocean is almost exclusively in the form of \ce{NO_{3}^{-}} , except where \ce{NO_{2}^{-}}  accumulates substantially within two depth horizons. The primary nitrite maximum, where nitrite concentrations reach hundreds of nanomolar to occasionally a few micromolar, is a global feature at the base of the photic zone, and formed by leaking algal cells \citep{Lomas2006} and/or an imbalance in the ammonium and nitrite oxidation steps of nitrification \citep{Santoro2013}. The secondary nitrite maximum is a broad feature in the oxygen deficient zones of the eastern tropical Pacific and Arabian Sea, where oxygen levels are reduced to $ <10$ nmol/L \citep{Revsbech2009}. These secondary nitrite maxima can achieve several micromolar in concentration (e.g., \citealt{Babbin2014}) and are indicative of active denitrification regimes.

Overall, previous work has concluded that high [\ce{NO_{X}^{-}}] (micromolar to millimolar) was present in the prebiotic ocean, under the assumption that the only sink of \ce{NO_{X}^{-}}~in the prebiotic ocean was processing at hydrothermal vents, and that atmospherically-supplied \ce{NO_{X}^{-}}~was otherwise stable in prebiotic waters. However, in the anoxic prebiotic environment, \ce{NO_2^-}~and \ce{NO_3^-}~are vulnerable to reduction to less soluble forms due to UV photochemistry and reactions with reductants (e.g., \ce{Fe^{2+}}). These reduced species (NO, N$_2$O, N$_2$) can then escape to the atmosphere, depleting the oceanic nitrogen pool \citep{Carpenter2015}. 

In this paper, we explore the impact of these chemical sinks on the predicted concentrations of \ce{NO_{X}^{-}}~in prebiotic waters. In Section~\ref{sec:kinetics}, we carry out a kinetic calculation, comparing the supply of \ce{NO_{X}^{-}}~to natural waters to the sinks of \ce{NO_{X}^{-}}~from reduction reactions and photochemistry to estimate the steady-state concentration of \ce{NO_{X}^{-}}~in oceans and ponds with T$=0-50^\circ$ C and pH$=6.5-8$. In Section~\ref{sec:eq}, we examine the thermochemical stability of \ce{NO_3^-}~and \ce{NO_2^-}~in prebiotic conditions, finding results consilient with our kinetic calculations. In Section~\ref{sec:disc}, we discuss our calculations and explore their implications for prebiotic \ce{NO_{X}^{-}}~~levels, postulated prebiotic chemistries,  and origin-of-life scenarios. We summarize our conclusions in Section~\ref{sec:conc}.

\section{Kinetic Steady-State\label{sec:kinetics}}

In this section, we calculate abiotic loss of aqueous \ce{NO_2^-}~and \ce{NO_3^-}~due to reduction by UV light and \ce{Fe^{2+}}, and compare to the \ce{NO_{X}^{-}}~supply rate from processes such as lightning fixation and exogenous delivery, to estimate steady-state \ce{NO_{X}^{-}}~concentrations. We focus on these processes because they are dominant under the range of conditions available on early Earth. If abiotic \ce{NO_{X}^{-}}~destruction is slow compared to abiotic \ce{NO_{X}^{-}}~production, then it is possible for high levels of \ce{NO_{X}^{-}}~to build up in natural waters. If destruction is fast compared to production, then \ce{NO_{X}^{-}}~levels will be low. 

\subsection{NO$_X^-$ Production \label{sec:kinetics_supply}}
On modern Earth, lightning fixation is the largest non-biological natural source of fixed nitrogen, via high-energy shocks which form free radicals and disrupt N$_2$ \citep{Mancinelli1988}. \citet{Wong2017} used  photochemical models calibrated with GCM results to model the supply of \ce{NO_{X}^{-}}~to the prebiotic surface due to lightning fixation. Most NO$_X^-$ reached the surface in the form of HNO, which in aqueous settings would then undergo dissociation, homologation, and decay reactions, ultimately yielding soluble \ce{NO_2^-}~and \ce{NO_3^-}, and gaseous N$_2$O \citep{Mancinelli1988, Summers2007}. \citet{Stanton2018} have recently demonstrated the reduction of NO to N$_2$O in ferrous waters; we may speculate similar chemodenitrification to occur with NO$^-$. It is difficult to calculate the fraction of fixed nitrogen that escapes back to the atmosphere as N$_2$O, because the kinetics of these transformations have not been quantified for NO$^-$. We follow \citet{Wong2017} in neglecting the sink to N$_2$O formation and assuming all NO$_X^-$ supplied to the surface from the atmosphere eventually yields \ce{NO_2^-}~and/or \ce{NO_3^-}; this means our calculations may overestimate NO$_X^-$ supply and hence [\ce{NO_{X}^{-}}]. \citet{Wong2017} report a dominant factor controlling the surface flux of NO$_X^-$ to be the partial pressure of CO$_2$ in the atmosphere; the minimum NO$_X^-$ flux, $\phi_{NO_{X}^{-}}$, was $\phi_{NO_{X}^{-}}=2.5\times10^5~\text{cm}^{-2}~\text{s}^{-1}$ for pCO$_2=0.1$ bar, and the maximum was $\phi_{NO_{X}^{-}}=6.5\times10^8~\text{cm}^{-2}~\text{s}^{-1}$ for pCO$_2=1$ bar. 

We used the lightning atmospheric chemistry model from \citet{Ardaseva2017} to explore the production of NO$_{X}^-$ by lightning as a function of pN$_2$, pCO$_2$, and lightning flash rate, to verify the upper bound in NO$_{X}^-$ production rate identified by \citet{Wong2017}. This model uses the freeze-out temperature approximation; a more sophisticated approach would involve shock modeling. We compare our lightning model calculations to the experimental results tabulated in \citet{Mvondo2001}, and find we reproduce these results well for CO$_2$ mixing ratios $\gtrsim 0.2$; at lower CO$_2$ concentrations, we overpredict NO$_{X}^-$ production. We therefore only apply our model to atmospheres with CO$_2$ mixing ratios $\geq 0.2$.

We find NO$_{X}^-$ production to be strong functions of pN$_2$/pCO$_2$ and pCO$_2$.  NO$_{X}^-$ production decreases as pN$_2$/pCO$_2$ increases, because the probability of N atoms recombining to N$_2$ is higher (as opposed to reacting with CO$_2$-derived oxygen to form NO$_{X}^-$). We find NO$_{X}^-$ production increases with pCO$_2$ over the range pCO$_2=0.1-10$ bar. This contrasts to \citet{Wong2017}, who report a maximum in NO$_{X}^-$ production at pCO$_2=1$ bar; this is because \citet{Wong2017} calculate lower lightning flash rates for high pCO$_2$ due to lack of moist convection in the warm troposphere they calculate for pCO$_2=10$ bar, while we fix the lightning flash rate. If we extrapolate the finding of \citet{Marty2013} that pCO$_2\leq0.7$ bar and pN$_2\geq0.5$ bar from 3-3.5 Ga to the prebiotic era (c. 3.9 Ga) and assume lightning energies and flash densities similar to modern Earth, we find a tropospheric NO$_{X}^-$ production rate of $\phi_{NO_{X}^{-}}<10^9~\text{cm}^{-2}~\text{s}^{-1}$, in concordance with \citet{Wong2017}. Biological fixation tends to decrease pN$_2$, suggesting that pN$_2$ was not lower in the prebiotic era than in the Archaean \citep{Johnson2017}. \citet{Krissansen-Totton2018} calculate that weathering restricted pCO$_2\leq 1$ bar at 4 Ga. We consequently retain  $6.5\times10^8~\text{cm}^{-2}~\text{s}^{-1}$ of \citet{Wong2017} as the upper bound on $\phi_{NO_{X}^{-}}$, but caution that if pCO$_2$ were higher or pN$_2$ lower than what we consider, $\phi_{NO_{X}^{-}}$ could have been up to an order of magnitude higher. For more details, see SI Section S4. 

Comet delivery and impact fixation should also have supplied fixed nitrogen on prebiotic Earth; however, these mechanisms are thought to have supplied $\phi_{NO_{X}^{-}}<2\times 10^{9}\text{cm}^{-2} \text{s}^{-1}$ and typically less, well within the range bracketed by lightning production \citep{Laneuville2018}. \citet{Airapetian2016} suggest that energetic protons from flares on the young Sun might also have powered nitrogen fixation and the supply of NO$_X^-$ to the surface; however, they do not quantify the magnitude of this supply. We consequently focus on the NO$_X^-$ production flux range defined by lightning production in the model of \citet{Wong2017} in our work ($2.5\times10^5 - 6.5\times10^8~\text{cm}^{-2}~\text{s}^{-1}$).

\subsection{NO$_X^-$ Destruction \label{sec:kinetics_loss}}
We consider three processes in calculating \ce{NO_{X}^{-}}~destruction in natural waters: processing at hydrothermal vents, UV photolysis, and reactions with \ce{Fe^{2+}}. Past work has focused on processing at vents; in this work, we consider the effects of UV and \ce{Fe^{2+}}~as well. SI Section S5 explores these processes in detail, along with other \ce{NO_{X}^{-}}~loss processes we neglected in this work because they are dominated by the processes we consider here. 

The presence of UV light on early Earth is attested to by the sulfur mass-independent fractionation (SMIF) signal \citep{Farquhar2001}. The presence of \ce{Fe^{2+}} is attested by the presence of banded iron formations (BIFs) and other geological evidence \citep{Cloud1973, Walker1985, Klein2005, Li2013}. Recent estimates place [\ce{Fe^{2+}}]$=30-600 \mu$M in early Archean oceans, with higher [\ce{Fe^{2+}}] in the aphotic deep oceans \citep{Tosca2016, Halevy2017, Konhauser2017, Zheng2018}. In this work, we explore [\ce{Fe^{2+}}]$=10-600 \mu$M, bracketing this range. We do not consider reactions with other reductants, such as Mn$^{2+}$, owing to paucity of constraints on the kinetics of these processes; consequently, we may underestimate \ce{NO_{X}^{-}}~reduction rates in prebiotic natural waters.

The rates of \ce{NO_{X}^{-}}~photolysis by UV and reduction by \ce{Fe^{2+}}~are dependent on temperature and pH. In this work, we consider $\text{pH}=6.5-8$  and $T=273-323$ K, motivated by modeling work which predicts the early ocean to have been circumneutral ($6.3\leq \text{pH} \leq 7.2$) and temperate ($271 \leq T \leq 314$ K) (\citet{Krissansen-Totton2018ocean}; see also \citet{Halevy2017}). We consider the sensitivity of our conclusions to these assumptions in Section~\ref{sec:kinetic_sensivity}.

\subsubsection{NO$_X^-$ Destruction in Vents}
The hot and acidic conditions at hydrothermal vents can destroy \ce{NO_{X}^{-}}~\citep{Ray1917,Brandes1998, Summers2005}. It is debated how extreme conditions need to be to consume \ce{NO_{X}^{-}}. Under the assumptions that \ce{NO_{X}^{-}}~is removed with unit efficiency at and only at black smoker-type vents ($T\lesssim 405^\circ$ C, pH$=1-2$, \citet{Martin2008}), \citet{Wong2017} propose \ce{NO_{X}^{-}}~destruction to be characterized by a first-order process with rate constant $k_{vents}=8\times10^{-17}~\text{s}^{-1}$. Under the assumption that circulation through any hydrothermal vent would destroy 100\% of \ce{NO_{X}^{-}}, \citet{Laneuville2018} instead propose $k_{vents}=1\times10^{-14}~\text{s}^{-1}$. We explore the range $k_{vents}=8\times10^{-17} - 1\times10^{-14}~\text{s}^{-1}$ in this work. 

\subsubsection{NO$_X^-$ Destruction by UV}
Irradiation by UV light in natural waters net photolyzes \ce{NO_3^-}~to \ce{NO_2^-}, and \ce{NO_2^-}~to NO, which escapes the ocean to the atmosphere or is reduced to N$_2$O \citep{Spokes1996, Mack1999,Fanning2000, Carpenter2015, Stanton2018}: 

\begin{align}
NO_3^- + h\nu \rightarrow NO_2^- + \frac{1}{2}O_2 \label{eqn:nitratephot}\\
NO_2^- + H_2O +  h\nu \rightarrow NO + OH + OH^- \label{eqn:nitritephot}
\end{align}

These processes are thought to be first order and have been measured both in the oceans and in lakes. In the modern surface ocean, these processes have median rate constants $k_{NO_{3}^{-}, h\nu} = 2.3\times10^{-8}~\text{s}^{-1}$ and $k_{NO_{2}^{-},h\nu}=1.2 \times 10^{-6}~\text{s}^{-1}$ for nitrate and nitrite, respectively \citep{Zafiriou1979nitrate, Zafiriou1979nitrite, Minero2007}. In pure water, \ce{NO_2^-}~reacts with OH to reform \ce{NO_3^-}; however, in the presence of OH scavengers like bicarbonate or Br$^-$, \ce{NO_2^-}~is lost with 20-100\% efficiency \citep{Treinin1970, Zafiriou1974, Zafiriou1979nitrite, Zafiriou1984}.

Nitrite and nitrite photolysis rates, as measured by OH production, depend modestly on temperature. At $T=0^\circ$ C, photolysis rates are $\geq0.5\times$ the rates at $T=25^\circ$ C, and at $T=50^\circ$ C, photolysis rates are $\leq 2\times$ the rates at $T=25^\circ$ C  \citep{Zellner1990, Mack1999}. The reaction rates measured by \citet{Zafiriou1979nitrite} and \citet{Zafiriou1979nitrate} were measured at ambient temperature. Under the assumption that these ambient conditions corresponded to $T\approx 25^\circ$ C, we explore  $k_{NO_{3}^{-}, h\nu} = 1.1-4.6\times10^{-8}~\text{s}^{-1}$ and $k_{NO_{2}^{-},h\nu}=0.6-2.4 \times 10^{-6}~\text{s}^{-1}$, to account for the variation in photolysis rates with temperature for $T=0-50^\circ$ C. 

We calculate the global rate of nitrate/nitrite photolysis following a procedure motivated by that of \citet{Zafiriou1979nitrate} and \citet{Zafiriou1979nitrite}. We assume that nitrate and nitrite are lost at rates equal to half their surface photolysis rates down to their photic depths, and that the loss rates are zero below this threshold. The modern photic depth for nitrate and nitrite at the equator are 5 m and 10 m, respectively; to average over latitude and obtain global mean photic depths, we scale these photic depths by $2/3$ \citep{Zafiriou1979nitrate, Zafiriou1979nitrite,Cronin2014}. In the modern ocean, intense consumption by phototrophic microbes depletes surficial \ce{NO_{X}^{-}}, meaning that \ce{NO_{X}^{-}}~photolysis is only significant in upwelling areas where \ce{NO_{X}^{-}}~is maintained at high concentrations due to supply from below. In the absence of biology, surface \ce{NO_{X}^{-}}~would not be depleted, and \ce{NO_{X}^{-}}~would be photolyzed from $100\%$ of the surface area of the ocean. When calculating the loss of \ce{NO_{X}^{-}}~in pond and lake environments, we assume the same photic depths, and continue to take \ce{NO_{X}^{-}}~to be lost from 100\% of the surface area. We neglect possible enhancements in the conversion rate due to factors such as availability of shorter-wavelength UV radiation on early Earth and the anoxic nature of prebiotic natural waters, and we assume the lowest proposed efficiency for net loss of \ce{NO_2^-}~to photolysis (i.e., 20\%). Consequently, our estimates of \ce{NO_{X}^{-}}~photolysis should be considered lower bounds.

\subsubsection{Reduction of \ce{NO_2^-}~by \ce{Fe^{2+}}~to Nitrogenous Gas}

\ce{Fe^{2+}}~reduces \ce{NO_2^-}~to yield nitrogenous gas \citep{Buchwald2016}:
\begin{align}
2NO_{2}^{-} + 4 Fe^{2+} + 5H_{2}O \rightarrow N_{2}O (g) + 4FeOOH + 6H^{+} \label{eqn:jones}\\
NO_{2}^{-} + 3 Fe^{2+} + 4H_{2}O \rightarrow N_{2} (g) + 3FeOOH + 5H^{+} \label{eqn:buchwald}
\end{align}

Recent kinetic studies of these reactions are consistent with first-order kinetics with respect to both reactants and second order kinetics overall, with rate constants that are dependent on pH \citep{Jones2015, Buchwald2016, Grabb2017}. From the data of \citet{Buchwald2016} and \citet{Grabb2017}, we extract $k_{NO_{2}^{-}, Fe^{2+}} = 3\times10^{-5} - 1\times10^{-2}~\text{M}^{-1}~\text{s}^{-1}$ over pH=6.5-8 and T$=25^\circ$ C (SI Section S5).  

$k_{NO_{2}^{-}, Fe^{2+}}$ depends on whether \ce{Fe^{2+}}~or \ce{NO_2^-}~is in excess, with reaction rates up to an order of magnitude lower if [\ce{NO_2^-}]$>$[\ce{Fe^{2+}}] compared to if [\ce{NO_2^-}]$<$[\ce{Fe^{2+}}] \citep{Jones2015}. To account for the potential dependence on relative \ce{NO_2^-}~and \ce{Fe^{2+}}~concentrations, we assign $k_{NO_{2}^{-}, Fe^{2+}}([\ce{NO_2^-}]>[\ce{Fe^{2+}}])=0.1 k_{NO_{2}^{-}, Fe^{2+}}([\ce{NO_2^-}]<[\ce{Fe^{2+}}])$. Further, The activation energy $E_A$ for \ce{NO_2^-}~reduction by dissolved \ce{Fe^{2+}}~has not been measured to our knowledge, but is $18.4~\text{kJ}~\text{mol}^{-1}$ for \ce{NO_2^-}~reduction by mineralized \ce{Fe^{2+}}, and is $70~\text{kJ}~\text{mol}^{-1}$  for \ce{NO_3^-}~reduction by dissolved \ce{Fe^{2+}}~\citep{Ottley1997, Samarkin2010}. Mineralized \ce{Fe^{2+}}~is a more effective reductant than dissolved \ce{Fe^{2+}}, and \ce{NO_2^-}~is more reactive than \ce{NO_3^-}, suggesting $18.4 \leq E_A \leq 70 ~\text{kJ}~\text{mol}^{-1}$. To ensure we do not underestimate the possible range of $k_{NO_{2}^{-}, Fe^{2+}}$, we take $E_A=70 ~\text{kJ}~\text{mol}^{-1}$.  

Combining these effects, in total we consider $k_{NO_{2}^{-}, Fe^{2+}}=2\times 10^{-6} - 9\times10^{-2}~\text{M}^{-1}~\text{s}^{-1}$ if [\ce{NO_2^-}]$<$[\ce{Fe^{2+}}] and $k_{NO_{2}^{-}, Fe^{2+}}=2\times 10^{-7} - 9\times10^{-3}~\text{M}^{-1}~\text{s}^{-1}$ if [\ce{NO_2^-}]$>$[\ce{Fe^{2+}}], corresponding to pH$=6.5-8$ and $T=0-50^\circ$ C.

\subsubsection{Reduction of \ce{NO_3^-}~by \ce{Fe^{2+}}~to Nitrogenous Gas}

 \ce{Fe^{2+}}~can reduce \ce{NO_3^-}~to nitrogenous gas, with proposed reactions \citep{Postma1990,Samarkin2010}:
\begin{align}
2NO_{3}^{-} + 12 Fe^{2+} + 11H_{2}O \rightarrow N_{2}O (g) + 4Fe_{3}O_{4} + 22H^{+} \label{eqn:samarkinrxn}\\
2NO_{3}^{-} + 10 Fe^{2+} + 14 H_{2}O \rightarrow N_{2} (g) + 10 FeOOH + 18 H^{+} \label{eqn:postmarxn}
\end{align}

The kinetics of uncatalyzed room-temperature reduction of \ce{NO_3^-}~by \ce{Fe^{2+}}~at room temperature are uncertain, because \ce{NO_3^-}~reduction is very slow under these conditions and hence difficult to characterize in the laboratory. \citet{Ottley1997} reports the detection of uncatalyzed \ce{NO_3^-}~reduction by \ce{Fe^{2+}}~at room temperature over timescales of a week. However, \citet{Picardal2012} report nondetections of \ce{NO_3^-}~reduction by \ce{Fe^{2+}}~ in sterile incubations carried out under conditions and timescales similar to those of \citet{Ottley1997}. As we are unable to favor one study above the other from available information, we consider a range of  $k_{NO_{3}^{-}, Fe^{2+}}=0 - 9\times10^{-4}~\text{M}^{-1}~\text{s}^{-1}$. The lower bound is derived from the reported nondetections of \citet{Picardal2012}. The upper bound is derived from the study of \citet{Ottley1997}, and corresponds to pH$=8$ and $T=50^\circ$ C, which should be the maximum rate possible over pH$=6.2-9$ and $T=0-50^\circ$ C \citep{Petersen1979}. We assume that the reduction of \ce{NO_3^-}~by dissolved \ce{Fe^{2+}}~is, like the reduction of \ce{NO_2^-}, first-order with respect to both reactants, for a second-order reaction overall, motivated by the generally similar kinetics of \ce{NO_2^-}~and \ce{NO_3^-}~reduction by Fe(0), H$_2$, and mineralized \ce{Fe^{2+}}~\citep{Samarkin2010,Zhu2012}. This range of $k_{NO_{3}^{-}, Fe^{2+}}$ is very large;  kinetic studies are required to constrain it.

\subsubsection{Other Reactions of \ce{NO_{X}^{-}}~with \ce{Fe^{2+}}}
\ce{Fe^{2+}}~can also reduce \ce{NO_2^-}~to NH$_3$, with proposed empirical reaction mechanism and rate law \citep{Summers1993}:

\begin{align}
NO_{2}^{-} + 6 Fe^{2+} + 7H^{+} \rightarrow 6 Fe^{III} + 2H_{2}O + NH_3 \label{eqn:nitritefento}\\
\frac{d[NH_3]}{dt} = k_{\ref{eqn:nitritefento}}[NO_{2}^{-} ][Fe^{2+}]^{1.8} (pH=7.9)
\end{align}

This reaction should be a negligible sink on [\ce{NO_2^-}] compared to \ce{Fe^{2+}}~reduction of \ce{NO_{X}^{-}}~to nitrogenous gas; to demonstrate this, we evaluate it for $k_{\ref{eqn:nitritefento}}=4.2\times10^{-5}$ s$^{-1}$ M$^{-1.8}$, corresponding to the maximum rate (pH$=7.6$, $T=40^\circ$C) measured by \citet{Summers1993}. UV photolysis and reduction by \ce{Fe^{2+}} dominate this process over the [\ce{NO_2^-}] range we consider here (Figure~\ref{fig:kinetics_ocean}).

Similarly, the anaerobic ammonium oxidation by \ce{NO_2^-}~(anammox) to N$_2$ is be negligible compared to other processes; to illustrate this, we compute its rate for [NH$_3$]$=6\times10^{-7}$ M and pH$=6.5$, $T=50^\circ~\text{C}$, corresponding to conditions maximize the reaction rate while conforming to the constraints \citet{Kasting1982} and \citet{Krissansen-Totton2018ocean}. We take the rate law from \citet{Nguyen2003} following \citet{Laneuville2018}:

\begin{align}
NO_2^- + NH_4^+ \rightarrow N_2 + 2H_2O \label{eqn:nitriteammonian}\\
\frac{d[N_2]}{dt} = A\exp(-E/RT) [NH_3] [HNO_2]^2
\end{align} 

With $A=\exp(37.8)~\text{M}^{-2}~\text{s}^{-1}= 2.6\times10^{16} ~\text{M}^{-2} ~\text{s}^{-1}$ and $E=65.7 \text{kJ}~\text{mol}^{-1}$. At $T=50^\circ~\text{C}$, this corresponds to a rate constant of $k_{\ref{eqn:nitriteammonian}}=A\exp(-E/RT)=6\times10^{5} \text{M}^{-2} \text{s}^{-1}$. UV photolysis and reduction by \ce{Fe^{2+}} dominate this process over the [\ce{NO_2^-}] range we consider here (Figure~\ref{fig:kinetics_ocean}).

 We note that the results of \citet{Nguyen2003} were based on experiments with reactant concentrations $\geq0.05$M. We assume this rate law to hold at lower concentrations as well; experimental studies are required to confirm this extrapolation. Our overall conclusions do not depend on these kinetics since these reactions are negligible compared to other processes.

\subsection{Calculation of [\ce{NO_{X}^{-}}] in Kinetic Steady-State}
We calculate the concentrations of \ce{NO_3^-}~and \ce{NO_2^-}~under the assumption of kinetic steady-state, i.e. that the loss rates of these molecules due to the destruction processes specified in Section~\ref{sec:kinetics_loss} and summarized in Table~\ref{tbl:loss_rxns} equals their supply from the atmosphere (Section~\ref{sec:kinetics_supply}).

\begin{sidewaystable}[H]
\begin{tabular} {p{5 cm}p{6cm} p{10.5 cm}}
 Process & Rate Law & Rate Constant\\
 \hline
 Hydrothermal Vents & $\frac{d[NO_X]}{dt}=k_{vents}[NO_X]$ & $k_{vents}=8\times10^{-17} - 1\times10^{-14}~\text{s}^{-1}$\\
  \ce{NO_2^-}~UV Photolysis & $\frac{d[NO_2^-]}{dt}=k_{NO_{2}^{-},h\nu}[NO_{2}^{-}]$ & $k_{NO_{2}^{-},h\nu}=0.6-2.4 \times 10^{-6}~\text{s}^{-1}$ \\ 
 \ce{NO_3^-}~UV Photolysis & $\frac{d[NO_3^-]}{dt}=k_{NO_{3}^{-},h\nu}[NO_{3}^{-}]$ & $k_{NO_{3}^{-}, h\nu} = 1.1-4.6\times10^{-8}~\text{s}^{-1}$ \\ 
 \ce{NO_2^-}~Reduction by \ce{Fe^{2+}}~to N$_2$, N$_2$O & $\frac{d[NO_2^-]}{dt} = k_{NO_{2}^{-}, Fe^{2+}} [Fe^{2+}] [NO_2^-]$ & $k_{NO_{2}^{-}, Fe^{2+}} = 2\times10^{-6} - 9\times10^{-2}~\text{M}^{-1}~\text{s}^{-1}$, ([\ce{Fe^{2+}}]$>$[\ce{NO_2^-}]); $k_{NO_{2}^{-}, Fe^{2+}} = 2\times10^{-7} - 9\times10^{-3}~\text{M}^{-1}~\text{s}^{-1}$, ([\ce{Fe^{2+}}]$<$[\ce{NO_2^-}])\\
  \ce{NO_3^-}~Reduction by \ce{Fe^{2+}}~to N$_2$, N$_2$O & $\frac{d[NO_3^-]}{dt} = k_{NO_{3}^{-}, Fe^{2+}}[NO_{3}^{-} ][Fe^{2+}]$ & $k_{NO_{3}^{-}, Fe^{2+}} = 0 - 9\times10^{-4}~\text{M}^{-1}~\text{s}^{-1}$  \\
  \ce{NO_2^-}~Reduction by \ce{Fe^{2+}}~to NH$_3$ & $\frac{d[NH_3]}{dt} = k_{\ref{eqn:nitritefento}}[NO_{2}^{-} ][Fe^{2+}]^{1.8}$ & $k_{\ref{eqn:nitritefento}}=4.2\times10^{-5}$ M$^{-1.8}$ s$^{-1}$  (Max.) \\
  Anammox of \ce{NO_2^-}~and NH$_3$ & $\frac{d[N_2]}{dt} = k_{\ref{eqn:nitriteammonian}} [NH_3] [HNO_2]^2$&  $k_{\ref{eqn:nitriteammonian}}= 6\times10^{5}~\text{M}^{-2}~\text{s}^{-1}$ ($T=50^\circ$C) \\

\end{tabular}
\caption{Summary of \ce{NO_{X}^{-}}~loss process kinetics}
\label{tbl:loss_rxns}
\end{sidewaystable}

\begin{figure}[H]
\centering
\includegraphics[width=.8\linewidth]{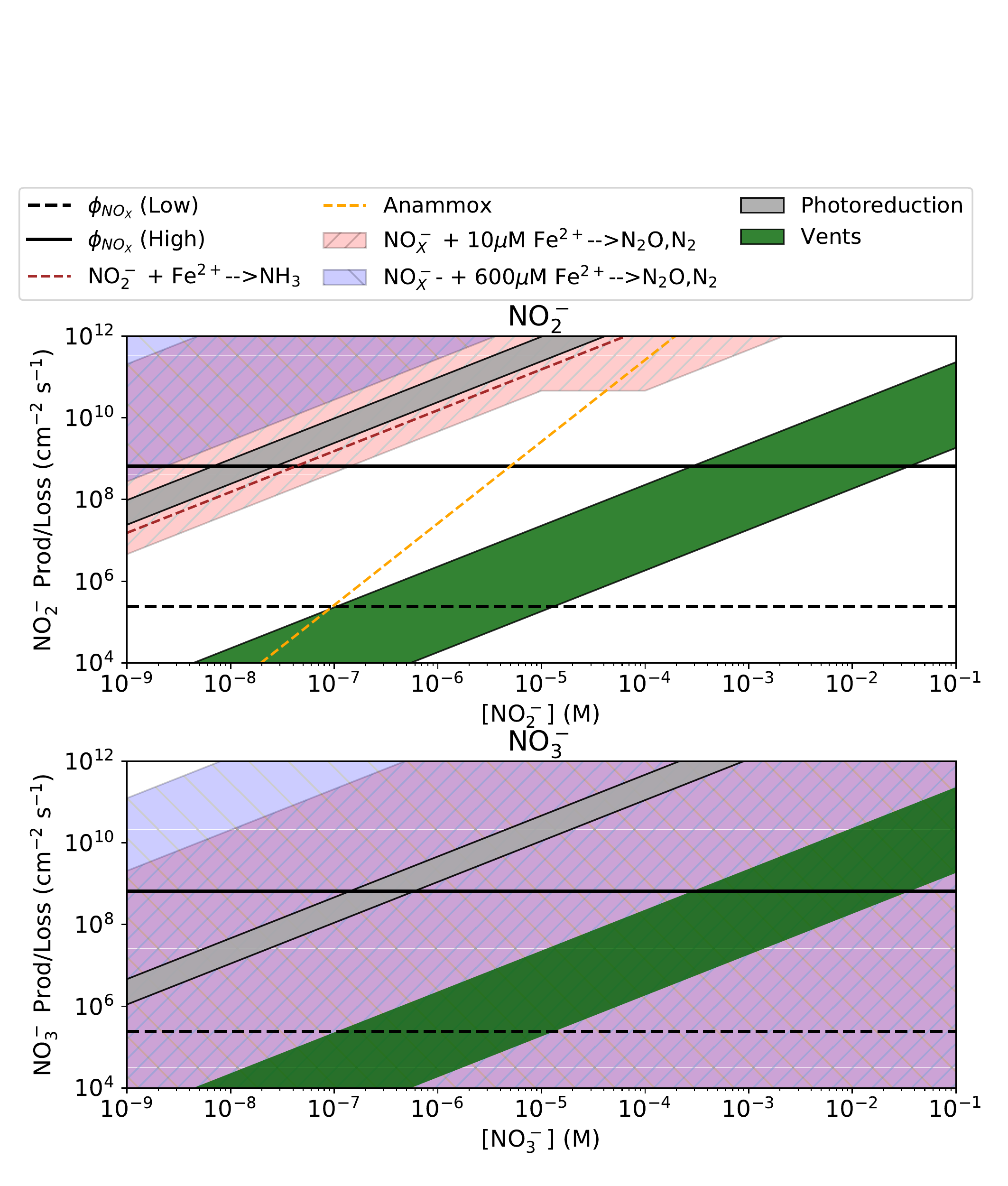}
\caption{Kinetic loss rates for oceanic \ce{NO_2^-}~(top) and \ce{NO_3^-}~(bottom) for the processes summarized in Table~\ref{tbl:loss_rxns}, as a function of [\ce{NO_2^-}] and [\ce{NO_3^-}]. Also plotted are the extremal $\phi_{NO_{X}^{-}}$ calculated by \citet{Wong2017}. The point at which the supply flux equals the destruction flux for a given process corresponds to the steady-state concentration for that process; the largest destruction flux (leftmost curve at given $\phi_{NO_{X}^{-}}$) dominates the system. Note that \ce{NO_3^-} reduction by \ce{Fe^{2+}} cannot be used to set an upper limit on [\ce{NO_3^-}], because we consider the possibility that $k_{NO_{3}^{-}, Fe^{2+}}=0$. \label{fig:kinetics_ocean}}
\end{figure}

To compare loss rates to the supply flux of \ce{NO_{X}^{-}}~~($\text{cm}^{-2}~\text{s}^{-1}$), we integrate over the water column, giving us a loss flux ($\text{cm}^{-2}~\text{s}^{-1}$). We consider both ocean and pond environments, corresponding to different families of postulated prebiotic chemistries (e.g., \citet{Patel2015} vs. \citet{Shibuya2016}). For the ocean, we adopt a depth of $d_{ocean}=3.8\times10^{5}$ cm, corresponding to the mean depth of the modern ocean \citep{CRC98}. Figure~\ref{fig:kinetics_ocean} presents the column-integrated destruction rates of oceanic \ce{NO_2^-}~and \ce{NO_3^-}~as functions of [\ce{NO_2^-}] and [\ce{NO_3^-}], as well as the range of plausible atmospheric supply rates from \citet{Wong2017}. The point at which the supply flux equals the destruction flux for a given process corresponds to the steady-state concentration for that process.

\begin{figure}[H]
\centering
\includegraphics[width=.8\linewidth]{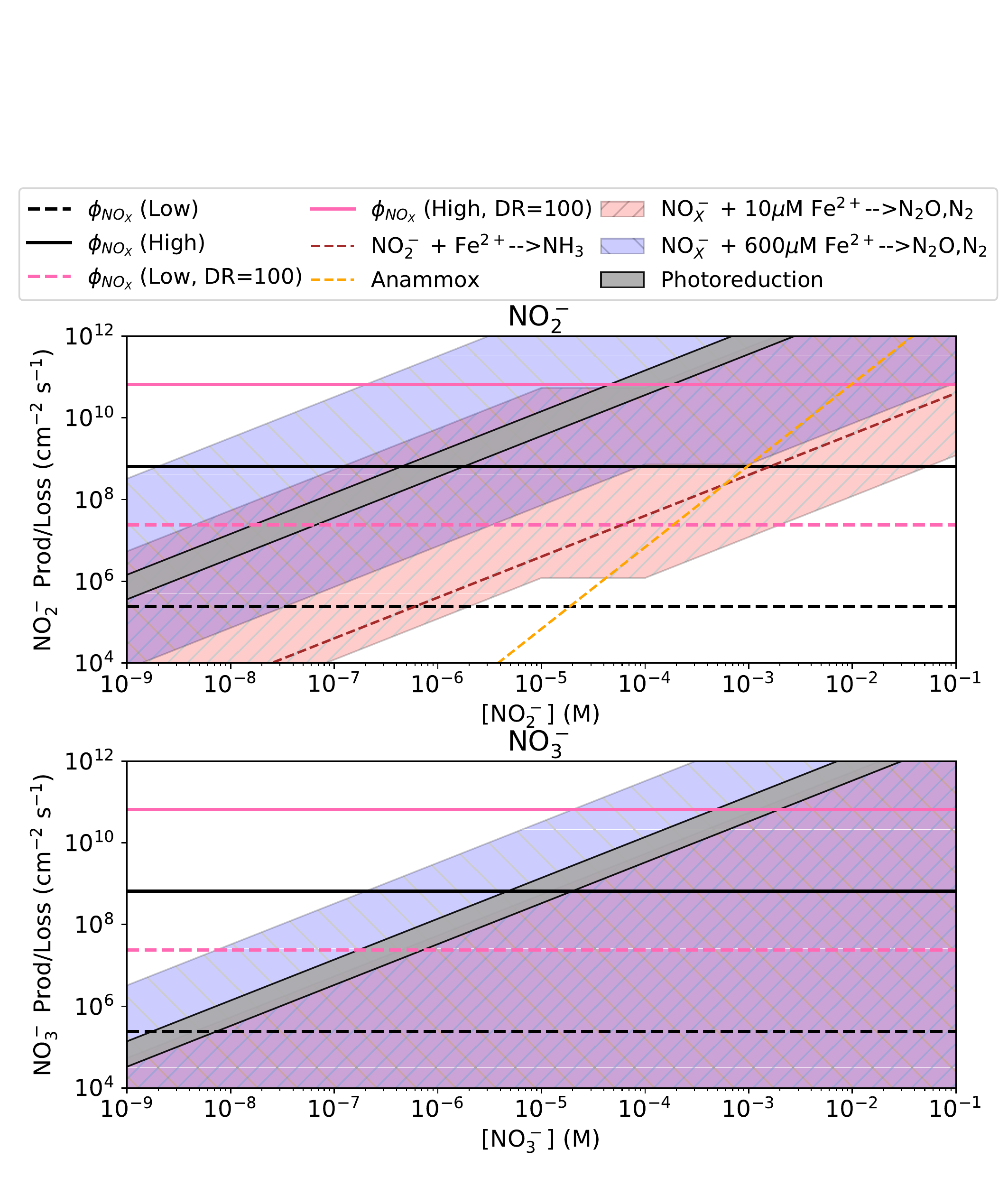}
\caption{Kinetic loss rates for pond \ce{NO_2^-}~(top) and \ce{NO_3^-}~(bottom) for the processes summarized in Table~\ref{tbl:loss_rxns}, as a function of [\ce{NO_2^-}] and [\ce{NO_3^-}]. Also plotted are the extremal $\phi_{NO_{X}^{-}}$ calculated by \citet{Wong2017}, as well as these $\phi_{NO_{X}^{-}}$ scaled by $100\times$ to simulate a lake/pond with a large DR and fast drainage. The point at which the supply flux equals the destruction flux for a given process corresponds to the steady-state concentration for that process; the largest destruction flux (leftmost curve at given $\phi_{NO_{X}^{-}}$) dominates the system. Note that \ce{NO_3^-} reduction by \ce{Fe^{2+}} cannot be used to set an upper limit on [\ce{NO_3^-}], because we consider the possibility that $k_{NO_{3}^{-}, Fe^{2+}}=0$. \label{fig:kinetics_pond}}
\end{figure}

For pond environments, a broad range of depths is possible. Larger depths correspond to lower [\ce{NO_{X}^{-}}], since more column is available over which \ce{NO_{X}^{-}}~is destroyed (or, equivalently, input \ce{NO_{X}^{-}}~flux is distributed over a larger column). To obtain an upper limit on plausible [\ce{NO_{X}^{-}}], we choose $d_{pond}=10$ cm, corresponding approximately to the summer depths of Don Juan Pond in Antarctica, which hosts millimolar abiotic \ce{NO_{X}^{-}}~\citep{Torii1977, Marion1997, Samarkin2010}. Figure~\ref{fig:kinetics_pond} presents the column-integrated destruction rate of pond \ce{NO_2^-}~and \ce{NO_3^-}~as a function of [\ce{NO_2^-}] and [\ce{NO_3^-}], as well as the range of plausible atmospheric supply rates from \citet{Wong2017}. Pond catchment areas can be much larger than their surface areas, meaning that ponds can concentrate atmospherically-delivered \ce{NO_{X}^{-}}~if the drainage timescale is short enough that the \ce{NO_{X}^{-}}~does not decay en route). The catchment area/surface area ratio is often termed the drainage ratio (DR). A study of catchment areas in southern England indicates means DR$=14$ for lakes and DR$=500$ for ponds \citep{Davies2008}. We therefore also present the supply fluxes scaled by a factor of 100, to simulate the potential concentrating effects of high DR. 

\begin{figure}[H]
\centering
\includegraphics[width=.8\linewidth]{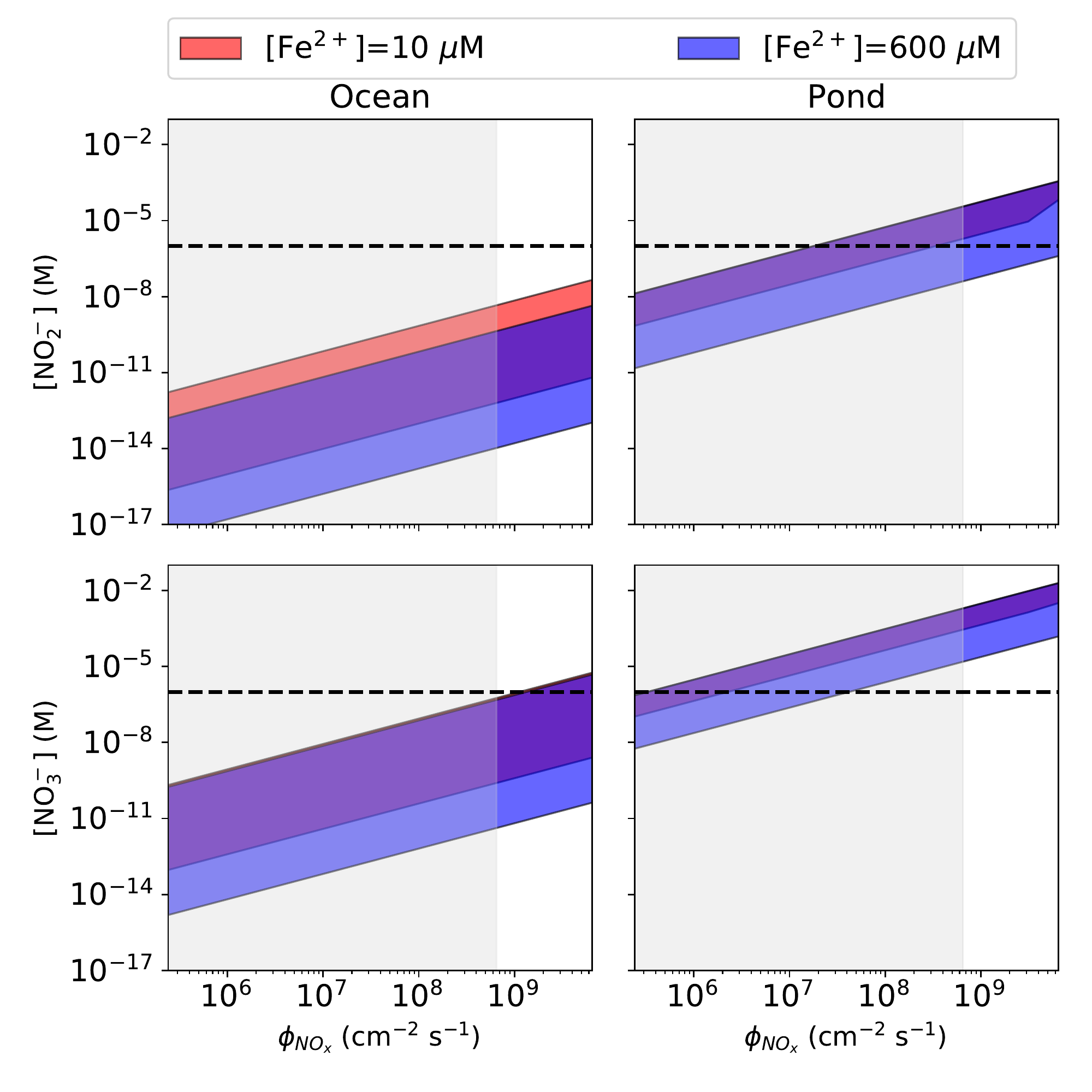}
\caption{Steady-state concentrations of \ce{NO_2^-}~and \ce{NO_3^-}~as a function of atmospheric supply flux with UV photolysis and reduction by \ce{Fe^{2+}}~to nitrogenous gas as the sinks, in the ocean and in a pond. The pond parameters ($d=10$ cm, DR$=100$) are favorable for \ce{NO_{X}^{-}}~accumulation, and hence should be considered an optimistic scenario. The horizontal dashed line demarcates micromolar concentrations, the putative boundary concentration for prebiotic relevance. The grey shaded area corresponds to the range of $\phi_{NO_{X}^{-}}$ calculated by \citet{Wong2017}. \label{fig:kinetics_eq_calc}}
\end{figure}

Figures~\ref{fig:kinetics_ocean} and ~\ref{fig:kinetics_pond} show that UV photolysis and reduction by \ce{Fe^{2+}}~to nitrogenous gas are the dominant sinks on \ce{NO_{X}^{-}}~in natural waters; at a given [\ce{NO_{X}^{-}}], the loss rates due to these processes are higher than the others, including processing at vents. We calculate [\ce{NO_2^-}] and [\ce{NO_3^-}] in the ocean as a function of $\phi_{NO_{X}^{-}}$ including UV photolysis and reduction by \ce{Fe^{2+}}~as sinks, and exploring the full range of reaction rate coefficients identified in Table~\ref{tbl:loss_rxns}. We assume the \ce{NO_{X}^{-}}~is supplied as 80\% \ce{NO_3^-}~and 20\% \ce{NO_2^-}, following the experimental work of \citet{Summers2007}; however, our results are not strongly sensitive to this assumption. We repeat this calculation for the case of a shallow pond with high drainage ratio and fast drainage ($d=10$ cm, DR$=100$), corresponding to a highly favorable scenario for \ce{NO_{X}^{-}}~accumulation. Hence, this should be considered an approximate upper bound on plausible \ce{NO_{X}^{-}}.  From this calculation, the upper bound on oceanic \ce{NO_{X}^{-}}~is $<10\mu$M, and typically $<1\mu$M across most of parameter space. Ponds with favorable drainage characteristics can accumulate much more \ce{NO_{X}^{-}}~(Figure~\ref{fig:kinetics_eq_calc}).

\subsection{Sensitivity to pH\label{sec:kinetic_sensivity}}
We have focused on pH$=6.5-8$, motivated by the findings of \citet{Halevy2017} and \citet{Krissansen-Totton2018ocean} that the early ocean was circumneutral. However, this finding is a model prediction. Additionally, lakes/ponds can be buffered by local factors to an even wider range of pH; on modern Earth, lake pH ranges from pH$<1$ to pH$>11$ \citep{Grant2000, Lohr2005}. Here, we consider the sensitivity of our results to our assumption of circumneutral pH.

As measured by OH production, \ce{NO_2^-}~photolysis rates vary by $\leq 2\times$ from pH$=4-11$, and nitrate photolysis rates by $\leq 3\times$ from pH$=2-14$, suggesting this process should be insensitive to pH \citep{Daniels1968, Zafiriou1987, Zellner1990}. However, for water with low concentrations of OH scavengers, nitrite photolysis should be reduced, since the OH can react with the photolysis products to reform the nitrite \citep{Zafiriou1974, Zafiriou1979nitrite}. Carbonate and bicarbonate are efficient OH scavengers. Consequently, achieving low OH scavenger concentrations requires pH$<6$, such that dissolved inorganic carbon is present primarily as CO$_2$ as opposed to bicarbonate or carbonate at higher pH. At such pH, \ce{NO_2^-}~is unstable. Hence, our choice of photolysis rate constants are valid from pH$=4-11$.

$k_{NO_{2}^{-}, Fe^{2+}}$  increases with pH for pH$=6-8.5$ \citep{Moraghan1977,Sorensen1991}. For pH$<6$, nitrite protonates and self-decomposes \citep{Ray1917,Park1988}. \citet{Brown1967} report fast reduction of \ce{NO_2^-}~by \ce{Fe^{2+}}~in alkaline solution \citep{Fanning2000} (though the experimental temperature is unclear). Thus, pH$\approx6$ probably represents a minimum on thermal loss of \ce{NO_2^-}.

\section{Thermochemical Equilibrium\label{sec:eq}}

In this section, we examine the stability of \ce{NO_3^-}~and \ce{NO_2^-}~in the anoxic early Earth environment under the assumption of thermal equilibrium. The purpose of this analysis is to test the implicit assumption of previous work that these molecules are stable in prebiotic waters absent processing at vents. Whether or not thermal equilibrium is achieved depends upon kinetic considerations. However, while our understanding of nitrogen kinetics in prebiotic natural waters may be incomplete (since we do not have an anoxic, prebiotic Earth-analog atmosphere-ocean system to study to confirm we have identified all relevant reactions), our equilibrium analysis depends only on known thermodynamic parameters, and hence is robust.

We consider the general speciation of nitrogen in a reducing atmosphere-ocean system as present on early Earth, with H$_2$ as our reductant (pH$_2\geq 1\times 10^{-3}$ bar on early Earth; \citet{Kasting2014}). We balance the nitrogen-converting redox half reactions with H$_2$ oxidation and take this to physically represent the amount of "reducing power" in the environment, even though the system could obtain its reducing power from other half reactions, e.g., Fe(II) oxidation. pN$_2$ on early Earth is known to have been comparable to present-day levels ($0.5<$pN$_2<1.1$ bar, \citet{Marty2013}). We consider an initial atmosphere-ocean system with atmospheric pN$_2=1.1$ bar, oceanic [N$_2$] in equilibrium with the atmosphere (i.e. saturated in N$_2$), and no other initial carrier of N. We assume an N$_2$-dominated atmosphere and ocean volume equal to modern. Then, the total inventory of nitrogen atoms $N_N$ in the atmosphere/ocean system is:

\begin{align}
N_{N} = 2\times ([N_2]V_{ocean} + (\frac{pN_{2}}{\mu g})(4\pi R_\text{Earth}^2))
\end{align}

Here, $V_{ocean}=1.4\times 10^{21}$ L is the volume of the ocean \citep{Pierazzo1999}, $\mu=28~\text{g/mol}$ is the mean molecular mass of the N$_2$-dominated atmosphere, and $g=981~\text{cm}~\text{s}^{-2}$ is the acceleration due to gravity. From Henry's Law, $[N_2] = H_{N_{2}} p_{N2}$, where $H_{N_{2}}=6.4\times 10^{-4}~\text{M}~\text{bar}^{-1}$ (Table~\ref{tbl:Henry}). Then, $N_{N}=4\times 10^{20}~\text{mol}$, comparable to present atmospheric N \citep{Johnson2015}.

We allow this N$_2$ to relax to equilibrium under a range of pH$_2$ and pH, and calculate the speciation of nitrogen compounds at equilibrium. To do this, we consider the possible reactions between the nitrogen species by balancing the individual half reactions for interconverting nitrogen species  with H$_2$ oxidation, and calculate cell potentials and logK for each reaction (see also SI, Section S2). We identified the reactions of each species with H2 with the largest logK; they are tabulated in Table~\ref{tbl:redoxapp}. The Gibbs free energies of formation used in this study, $\Delta G^\circ_{f}$, are compiled in Table~\ref{tbl:dG}.

\begin{table}[H]
\begin{tabular} {c c c}
\textbf{Species} & \textbf{$\Delta G^\circ_f$ (kJ/mol)} & Reference\\
\hline
NO$_3^-$ (aq) & -111.3 & \citet{CRC98}\\ 
NO$_2^-$ (aq) & -32.2 & \citet{CRC98} \\ 
NH$_4^+$ (aq) & -79.5 & \citet{CRC98} \\
NO (g) & 87.6 & \citet{CRC98}\\
N$_2$O (g) & 103.7 & \citet{CRC98} \\
N$_2$ (g) & 0 & \citet{CRC98}\\
N$_2$ (aq) & 18.8 & \citet{Amend2001}\\
O$_2$ (g) & 0 & \citet{CRC98}\\
O$_2$ (aq) & 16.54 & \citet{Amend2001}\\
H$_2$ (g) & 0 & \citet{CRC98}\\
H$_2$ (aq) & 17.72 & \citet{Amend2001}\\
H$_2$O & -237.1 & \citet{CRC98} \\
H$^+$ & 0 & \citet{CRC98} \\
\hline
\end{tabular}
\caption{Gibbs free energies of formation under standard conditions ($\Delta G^\circ_f$) for the species considered in this work.}
\label{tbl:dG}
\end{table}

\begin{table}[H]
\begin{tabular}{l l c c c}
\textbf{Number} & \textbf{Reaction} & \textbf{ $\Delta G^\circ_{rxn}$ (kJ/mol)} & \textbf{$E^\circ_{cell} (V)$} & \textbf{log K}\\
\hline
1 & $2NO_3^- + 12H^+ + 10e^- \rightarrow N_2 + 6H_2O$ & & & \\
 & $ 2NO_3^- + 2H^+ + 5H_2 \rightarrow N_2 + 6H_2O $ & -1202 & 1.25 & 210.4 \\
2 & $2NO_2^- + 8H^+ + 6e^- \rightarrow N_2 + 4H_2O$ & & & \\
 & $ 2NO_2^- + 2H^+ + 3H_2 \rightarrow N_2 + 4H_2O $ & -879.6 & 1.52 & 154.0 \\
3 & $NO + 6H^+ + 5e^- \rightarrow NH_4^+ + H_2O$ & & & \\
 & $ 2NO + 2H^+ + 5H_2 \rightarrow 2NH_4^+ + 2H_2O $ & -403.3 & 0.84 & 141.9 \\
4 & $N_2O + 10H^+ + 8e^- \rightarrow 2NH_4^+ + H_2O$ & & & \\
 & $ N_2O + 2H^+ + 4H_2 \rightarrow 2NH_4^+ + H_2O $ & -499.7 & 0.65 & 87.5\\
5 & $N_2 + 8H^+ + 6e^- \rightarrow 2NH_4^+ $ & & & \\
 & $ N_2 + 2H^+ + 3H_2 \rightarrow 2NH_4^+ $ & -159.0 & 0.27 & 27.8 \\
\end{tabular}
\caption{Half reactions and full cell reactions (balanced with $H_2$ oxidation) with the maximum logK's for the nitrogen species considered in this study. $\Delta G^\circ_{rxn}, E^\circ_{cell},$ and log K were calculated using the standard expressions (SI Section S2)}
\label{tbl:redoxapp}
\end{table}

We use these reactions to set up equations for concentrations at equilibrium using the definition of the equilibrium constant:

\begin{align}
K_1 = \frac{[N_2]}{[NO_3^-]^2 [H^+]^2 [H_2]^5}
\end{align}
\begin{align}
K_2 = \frac{[N_2]}{[NO_2^-]^2 [H^+]^2 [H_2]^3}
\end{align}
\begin{align}
K_{3} = \frac{[NH_4^+]^2}{[NO]^2 [H^+]^2 [H_2]^5}
\end{align}
\begin{align}
K_{4} = \frac{[NH_4^+]^2}{[N_{2}O] [H^+]^2 [H_2]^4}
\end{align}
\begin{align}
K_{5} = \frac{[NH_4^+]^2}{[N_2] [H^+]^2 [H_2]^3}
\end{align}

NO$_3^-$, NO$_2^-$, and NH$_4^+$ undergo further acid/base equilibration, with partitioning governed by the reaction pKa's (Table~\ref{tbl:AcidBase}):

\begin{align}
HA \rightarrow H^+ + A^-\\
K_a = \frac{[H^+][A^-]}{[HA]} = 10^{-pKa} 
\end{align}

Where $A^-$ is the acid and $HA$ its conjugate base.

\begin{table}[H]
\begin{tabular} {l l c}
\textbf{Reaction} & \textbf{pKa} & Reference\\
 \hline
$HNO_3 \leftrightarrow H^+ + NO_3^- $ & -1.38 & \citet{Dean1985}\\
$HNO_2 \leftrightarrow H^+ + NO_2^- $ & 3.25 & \citet{CRC98}\\
$NH_4^+ \leftrightarrow H^ + + NH_3 $ & 9.25 & \citet{CRC98} \\
\end{tabular}
\caption{pKa's for relevant dissociations of acids/bases in the redox network to their corresponding conjugate base/acid. }
\label{tbl:AcidBase}
\end{table}

Aqueous HNO$_3$, HNO$_2$, NO, N$_2$O, N$_2$, NH$_3$ exist in equilibrium with their gaseous forms, with partitioning specified by Henry's Law (Table~\ref{tbl:Henry}; \citet{Sander2015}):

\begin{align}
[X] = H_{X} \text{pX}
\end{align}

Where $H_X$ is the Henry's Law constant for species $X$, and pX its partial pressure.

\begin{table}[H]
\begin{tabular}{ l c c} 
\textbf{Species} & \textbf{$H$} (M/bar) & Reference \\
 \hline
HNO$_3$ & $2.6 \times 10^{6}$ & \citet{Chameides1984} \\
HNO$_2$ & $5\times 10^{1}$ & \citet{Chameides1984}  \\
NO & $1.9 \times 10^{-3}$ & \citet{Schwartz1981} \\
N$_2$O & $2.4 \times 10^{-2}$ & \citet{Sander2015} \\
N$_2$ & $6.4 \times 10^{-4}$ & \citet{Sander2015} \\ 
NH$_3$ & $6\times 10^{1}$ & \citet{Kavanaugh1980} \\
O$_2$ (g) & $1.3\times10^{-3}$ & \citet{Sander2015} \\
H$_2$ (g) & $7.8\times10^{-4}$ & \citet{Sander2015} \\
 \hline
\end{tabular}
\caption{Henry's Law constants ($H$) under standard conditions and zero salinity for the species considered in this analysis in aqueous solution. }
\label{tbl:Henry}
\end{table}

Finally, we have the mass balance constraint that the sum of nitrogen atoms across all species equals the initial nitrogen inventory $N_{N}$:

\begin{align}
N_{N} = ([NO_3^-]+[NO_2^-]+[NO]+2[N_2O]+2[N_2]+[NH_4^+]+[HNO_3]+[HNO_2]+[NH_3]) V_{ocean}\\ + \frac{pHNO_3+pHNO_2+pNO+2pN_2O+2pN_2+pNH_3}{\mu g}(4\pi R_\text{Earth}^2)
\end{align}

\begin{figure}
\includegraphics[scale=.35]{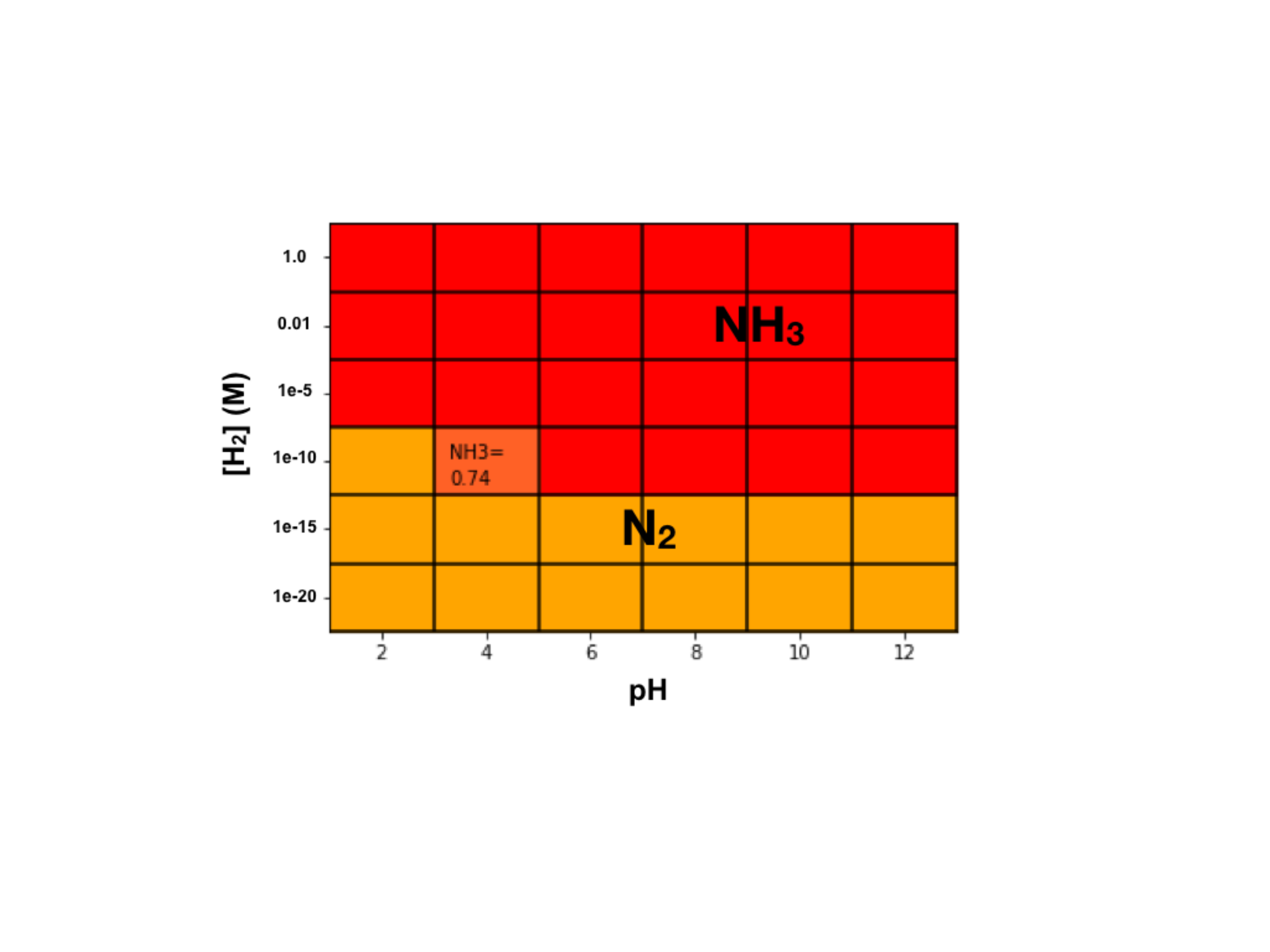}
\caption{Favored nitrogen species as a function of pH and hydrogen concentration. At high reducing powers ( $\geq10^{-5}~\text{M}~\text{H}_2$), at thermodynamic equilibrium, nitrogen will go to the (-3) oxidation state, i.e. NH$_3$ or NH$_{4}^{+}$. Below this reducing power, the favored state of nitrogen is N$_2$, which partitions mainly to the atmosphere. At pH$=4$ and [H$_2$]$=10^{-10}~\text{M}$, N$_2$ and NH$_3$ coexist. Lower H$_2$ concentrations are disfavored in conventional models of early Earth and not included in the analysis \citep{Kasting1993}.}
\label{fig:redoxeq}
\end{figure}

Taken together, this system provides us with 15 equations, 15 unknowns ([NO$_3^-$], [NO$_2^-$], [NO], [N$_2$O], [N$_2$], [NH$_4^+$]; [HNO$_3$], [HNO$_2$], [NH$_3$]; pHNO$_3$, pHNO$_2$, pNO, pN$_2$O, pN$_2$, pNH$_3$ ), and 3 prescribed conditions ([H$^+$], [H$_2$], $N_{N}$). We solve this system of equations for a range of pH ([H$^+$]) and reducing powers ([H$_2$]) for $N_{N}=4\times10^{20} \text{mol}$. We compute the fraction of atoms of N stored in each species (Figure~\ref{fig:redoxeq}).

Figure \ref{fig:redoxeq} shows the speciation of nitrogen by redox state. The two oxidation states of nitrogen favored under plausible hydrogen concentrations are (-3) and (0), corresponding to NH$_3$/NH$_{4}^{+}$ and N$_2$, respectively. N$_2$, which is favored at [H$_2$]$\leq 10^{5}$M, will partition mostly into the atmosphere. The distribution of NH$_{4}^{+}$, NH$_3$ (aq), and NH$_3$ (g) depends on the pH of the aqueous solution. The concentration of \ce{NO_3^-}~and \ce{NO_2^-}~is sub-picomolar over the scenarios considered here. These findings are insensitive to variations in pK, pKa, and $H_X$ from $T=2-45^\circ$ C, and variations in $H_X$ due to salinity from [NaCl]$=0-1$M; see SI Section S3 for details.

Overall, nitrate and nitrite are thermally unstable in the reducing conditions available on early Earth. If allowed to reach equilibrium, under most conditions \ce{NO_{X}^{-}}~will relax to a gaseous species like N$_2$. This result is consistent with \citet{VanCleemput1978}, who concluded that \ce{NO_2^-}~in anoxic soils should decay, generally to N$_2$.  The situation on reducing prebiotic Earth is very different from the situation on oxidizing modern Earth, where in equilibrium NO$_X^-$ is is thermodynamically favored \citep{Krissansen-totton2016}. Our thermochemical analysis confirms our kinetic analysis that previous work has overestimated prebiotic [NO$_X^-$], increasing our confidence in this conclusion.

\section{Discussion \label{sec:disc}}

\subsection{[\ce{NO_{X}^{-}}] in the Prebiotic Oceans}
Previous work has concluded that [\ce{NO_{X}^{-}}] in the prebiotic oceans was high, on the assumption that the dominant sink of \ce{NO_{X}^{-}}~in the prebiotic ocean was processing at hydrothermal vents and that it was otherwise stable in the ocean  \citep{Wong2017,Laneuville2018}. However, \ce{NO_{X}^{-}}~is thermodynamically unstable in reducing environments. UV photolysis and reduction by \ce{Fe^{2+}}~are much stronger sinks than processing at vents, and restrict \ce{NO_{X}^{-}}~to orders of magnitude lower concentrations than previously suggested. Reduction by \ce{Fe^{2+}}~and UV photolysis suppresses \ce{NO_{X}^{-}}~to sub-micromolar concentrations across most of the plausible parameter space. $>1~\mu$M \ce{NO_{X}^{-}}~requires $\phi_{NO_{X}^{-}}>1\times10^{9}~\text{cm}^{-2}~\text{s}^{-1}$, higher than has been suggested in the literature. Achieving such high $\phi_{NO_{X}^{-}}$ requires some combination of a high lightning flash rate, high pCO$_2$, and low pN$_2$/pCO$_2$. The required pCO$_2$ and pN$_2$ are not favored by the available geochemical evidence, and we lack a robust prescription for global lightning flash rates, leading us to disfavor this possibility. Consequently, prebiotic oceanic \ce{NO_{X}^{-}}~was likely sub-micromolar.

\citet{Wong2017} point out that \citet{Fitzsimmons2014} have detected dissolved iron thousands of kilometers from hydrothermal sources, and suggest that the survival of this iron for such large distances on \ce{NO_{X}^{-}}-rich modern Earth means that \ce{Fe^{2+}}~oxidation, by \ce{NO_{X}^{-}}~and other oxidants is inefficient. However, \citet{Fitzsimmons2014} also point out that only $0.02-1\%$ of hydrothermal Fe survives transport over these distances in the dissolved phase, meaning that the vast majority of hydrothermal Fe is oxidized. Moreover, the Fe that does avoid oxidation is thought to do so by forming colloids and/or by complexing with organic ligands \citep{Sander2011vents, Hawkes2013, Fitzsimmons2014, Tagliabue2014}. In other words, hydrothermal Fe appears to survive long-distance transport because it is protected by complexing, mineralization, and colloidation, not because its reactions with oxidants are inefficient. Additionally, mineralized \ce{Fe^{2+}}~is typically a more effective reductant than dissolved \ce{Fe^{2+}}~\citep{Sorensen1991, Hansen1996, Dhakal2013}. We consequently argue that it is not possible to dismiss reduction by \ce{Fe^{2+}}~as a sink on \ce{NO_{X}^{-}}, especially in light of evidence that \ce{Fe^{2+}}~levels were high on early Earth.

\subsection{[\ce{NO_{X}^{-}}] in Prebiotic Ponds}
As in the oceans, reduction by \ce{Fe^{2+}}~and photochemical loss are major sinks of \ce{NO_{X}^{-}}. However, since ponds are much shallower than oceans, the impact of thermal reactions is muted, and UV photolysis is proportionated more important. 

 [\ce{NO_{X}^{-}}] could have been above oceanic in shallow ponds ($d\lesssim3$ m) with large DR and short transit times. Shallow ponds permit higher \ce{NO_{X}^{-}}~buildup because \ce{NO_{X}^{-}}~destruction processes have a shorter column over which to operate. Large drainage ratios permit ponds to collect \ce{NO_{X}^{-}}~rainout from a larger area. Short transit times minimize the probability the \ce{NO_{X}^{-}}~will decay en route due to encounters with reductants in the soil. For a pond with $d=10$ cm, DR$=100$, and fast drainage, [\ce{NO_{X}^{-}}] can build up to micromolar concentrations for $\phi_{NO_{X}^{-}}\geq 4\times10^{7}~\text{cm}^{-2}~\text{s}^{-1}$, and at lower $\phi_{NO_{X}^{-}}$ if the pond is cold and acidic. For $\phi_{NO_{X}^{-}}= 6.5\times10^{8}~\text{cm}^{-2}~\text{s}^{-1}$, [\ce{NO_{X}^{-}}] can build up to near-millimolar concentrations in such a lake. 

We consider whether such a pond is plausible. A study of water bodies in southern England indicated that the ratio between the total catchment area and total surface area for lakes and ponds was 14 and 500, respectively, and a study of boreal lakes in Sweden found DR as high as 1000, indicating DR=100 to be plausible \citep{Sobek2003,Davies2008}. Assuming neutral, room-temperature groundwaters with [\ce{Fe^{2+}}]$ \leq 10^{-4}$ M, the lifetime of \ce{NO_{X}^{-}}~is $\geq 400 $ days, implying transit times $\leq 400$ days are required to ensure negligible decay of \ce{NO_{X}^{-}}~during transport. Catchment transit times $\leq$ 400 days exist, particularly for smaller catchments, but are not universal, indicating that only a subset of ponds will meet this criterion \citep{Rodhe1996, Brosig2008, Broxton2009,McGuire2006}. Further, if present, mineralized \ce{Fe^{2+}}~in the ground may more efficiently reduce \ce{NO_{X}^{-}}~\citep{Dhakal2013}. These challenges will be avoided in terrain where rain is immediately lost to the pond as surface runoff; this is especially likely to occur in catchments in bare, rocky terrain \citep{Li2011}. Ponds with high concentrations of \ce{NO_{X}^{-}} do exist on modern Earth, as predicted from our modeling; an example is Don Juan Pond, which is thought to be abiotic and which features [\ce{NO_{X}^{-}}] $=6$ mM \citep{Samarkin2010}. In summary, ponds with high [\ce{NO_{X}^{-}}] should have existed on early Earth, but were probably not typical; hence, pond prebiotic chemistries which require high \ce{NO_{X}^{-}}~must specify such a pond as part of their scenario.

\subsection{Implications for Prebiotic Chemistry}

Oceanic \ce{NO_{X}^{-}}~could only have achieved prebiotically relevant levels if atmospheric supply rates were very high. Achieving [\ce{NO_{X}^{-}}]$\geq1~\mu$M requires $\phi_{NO_{X}^{-}}\geq\times 10^9~\text{cm}^{-2}~\text{s}^{-1}$, which requires some combination of high flash rates, high pCO$_2$, and low pN$_2$/pCO$_2$. These conditions are not at present favored in the literature (\citet{Marty2013, Wong2017,Krissansen-Totton2018ocean}). Consequently, oceanic \ce{NO_{X}^{-}}-dependent origin-of-life scenarios (e.g., those that invoke \ce{NO_{X}^{-}}~as electron acceptors at deep-sea hydrothermal vents; \citet{Ducluzeau2009,Nitschke2013,Shibuya2016}) must invoke either extreme planetary parameters, or local circumstances which can concentrate [\ce{NO_{X}^{-}}] levels beyond the oceanic mean. 

\ce{NO_{X}^{-}}~could have achieved above-oceanic concentrations in favorable pond environments, i.e. ponds with large DR and short catchment transit times. Low temperatures and acidic pH would also favor \ce{NO_{X}^{-}}~buildup, especially as \ce{NO_3^-}. [\ce{NO_{X}^{-}}] could be even higher at polar latitudes where photolysis rates are suppressed by low UV surface radiances due to larger solar zenith angles \citep{Ranjan2017a}. Such a pond would be able to sustain [\ce{NO_{X}^{-}}]$\geq 1 \mu$M across most of the range of $\phi_{NO_{X}^{-}}$ calculated by \citet{Wong2017}. Such ponds are plausible but not typical, and hence must be explicitly invoked when considering \ce{NO_{X}^{-}}-dependent pond prebiotic chemistries. Their non-universality must also be considered when estimating the plausibility of \ce{NO_{X}^{-}}-dependent prebiotic chemistries.

\ce{NO_3^-}~is orders of magnitude more stable than \ce{NO_2^-}. This suggests that in both lake and oceanic environments, prebiotic \ce{NO_{X}^{-}}~should have existed primarily as \ce{NO_3^-}, as in natural waters on modern Earth and in experimental studies of abiotic nitrogen fixation \citep{Summers2007}. Consequently, \ce{NO_3^-}-utilizing prebiotic chemistries are more plausible than \ce{NO_2^-}-dependent prebiotic chemistries, and prebiotic chemists should consider using \ce{NO_3^-}~instead of \ce{NO_2^-}~in their studies.

In this work, we have focused on concentrations of \ce{NO_2^-}~and \ce{NO_3^-}, under the broad category of \ce{NO_{X}^{-}}. We have ignored more complex derivatives of these compounds. For example, \citet{Mariani2018} point out that under UV irradiation, \ce{NO_3^-}, Fe, and HCN combine to yield nitroprusside, a compound in which \ce{NO_{X}^{-}}~is protected from reduction by \ce{Fe^{2+}}~and which is stable in the dark on a timescale of $\geq5$ months. However, nitroprusside is unstable to irradiation by the visible light which accompanies UV irradiation \citep{Vesey1977, VanLoenen1979, Schulz2010}. Measurements of the kinetics of nitroprusside formation and destruction are required to determine the range of plausible steady-state concentrations of nitroprusside in prebiotic natural waters.

We note in passing that the prospects for abiotic \ce{NO_{X}^{-}}~buildup may be enhanced on planets orbiting M-dwarfs, due to their much lower surface UV irradiation and consequently much slower \ce{NO_{X}^{-}}~photolysis rate \citep{Ranjan2017c}. Consequently, \ce{NO_{X}^{-}}-dependent prebiotic chemistries may proceed especially well on such worlds relative to early Earth.

\subsection{Validity of Simplifying Assumptions}

In this work, we have approximated the activity of ionic species (e.g., \ce{NO_3^-}, H$^+$) by their concentrations, neglecting the effects of ion-ion and ion-water interactions on their reactivity. For the species relevant to this work, the activity coefficient $\gamma_C\geq0.26$ for solutions with ionic strengths $I\leq1$ (SI Section S1). For context, the ionic strength of the modern oceans is $I=0.7$, and studies of fluid inclusions in quartz suggests that Archean ocean salinity was $\lesssim$ modern \citep{CRC98, Marty2018}). Our order-of-magnitude conclusions are insensitive to such variations, motivating this simplifying assumption.

\citet{VanCleemput1983} suggest that the kinetics of \ce{NO_2^-}~reduction by \ce{Fe^{2+}}~are second-order with respect to nitrite concentrations at acidic pH. We repeated our analysis assuming second-order dependence on [\ce{NO_2^-}]; our conclusions were unchanged, indicating our analysis is insensitive to this possibility.

Our photolysis calculations assume photolysis rate constants equal to the modern value. While shortwave surface UV irradiation ($200-300$ nm) on anoxic early Earth was much higher than on modern Earth, surface UV irradiation over the full UV range ($200-400$ nm) was 20\% lower on prebiotic Earth compared to modern Earth, suggesting we may slightly overestimate the photolysis rate \citep{Ranjan2017a}. However, (1) our conclusions are robust to variations in photolysis rate of a few tens of percent, and (2) the magnitude of \ce{NO_{X}^{-}}~photolysis is sensitive to the action spectrum of \ce{NO_{X}^{-}}~photolysis; if shorter wavelengths are much more effective at photolyzing \ce{NO_{X}^{-}}, then our methods may underestimate photolysis rates \citep{Cockell2000UVhist, Claire2012, Ranjan2016}. Further measurements of the action spectrum of \ce{NO_{X}^{-}}~photolysis are required to rule on this possibility.

Our calculations assumes all NO$_X^-$ entering the ocean goes to \ce{NO_{X}^{-}}~and  neglects reactions of \ce{NO_{X}^{-}}~with other reductants which may have been abundant on early Earth, such as H$_2$, CH$_4$, or Mn$^{2+}$ \citep{Tian2005, Fischer2016, Zahnle2018}. Consequently, our estimates should be considered upper bounds on prebiotic [\ce{NO_{X}^{-}}]. 

This box-model approach we have taken averages over the entire natural water body under consideration. This approach permits us to place bounds on the mean concentrations of \ce{NO_{X}^{-}}~in prebiotic natural waters with minimal assumptions, and is in line with past work (e.g., \citet{Wong2017,Laneuville2018}). This approach is a good approximation to well-mixed shallow lakes and ponds. However, the oceans are not necessarily well-mixed; [\ce{NO_{X}^{-}}] may be a function of depth. Resolved, 1D models are required to probe this effect; mean oceanic concentration should be similar, but \ce{NO_{X}^{-}}~concentrations should be higher at the surface where it is supplied and lower at depth. In summary, our approach suffices for \ce{NO_{X}^{-}}~estimates in ponds, and for estimates of the mean \ce{NO_{X}^{-}}~concentration in the ocean, but resolving the heterogeneity of the ocean requires higher-dimensional models.

\subsection{Importance of Better Kinetic Constraints}
Measurements of \ce{NO_{X}^{-}}~kinetics under conditions relevant to the early Earth are scarce. While our calculations are motivated by and consistent with available data, improved measurements of these kinetics can decrease the uncertainty in these calculations and improve the confidence of these results. In particular: (1) the literature contains contradictory reports as to whether uncatalyzed \ce{NO_3^-}~reduction by \ce{Fe^{2+}}~is significant at room temperature \citep{Ottley1997, Picardal2012}. Experimental studies are required to resolve the dichotomy between these studies; if this process is indeed significant, as \citet{Ottley1997} report, then oceanic \ce{NO_3^-}~concentrations would be suppressed to concentrations lower than we model here. (2) The activation energy for reduction of \ce{NO_2^-}~by \ce{Fe^{2+}}~is not known; knowledge of this quantity would enable tighter constraints on $k_{NO_{2}^{-}, Fe^{2+}}>0$. (3) The rate constants for \ce{Fe^{2+}}~reduction used in this work are generally derived from measurements made at larger [\ce{Fe^{2+}}] than thought to have been available on early Earth. Determination of these rate constants at prebiotically plausible [\ce{Fe^{2+}}] ($10-600\mu$M) under early Earth conditions (e.g., anoxia) could confirm the applicability of these rate constants at lower [\ce{Fe^{2+}}].  The extension of studies like \citet{Stanton2018} for NO kinematics to NO$_X^-$ kinematics could improve the precision and potentially accuracy of this work. (4) Measuring the rate constant of \ce{NO_{X}^{-}}~photolysis in simulated prebiotic natural waters, under irradiation by a source simulating the prebiotic UV environment in both magnitude and spectral shape, could directly verify our extrapolation from modern photolysis rates, and in particular could confirm whether the shorter-wavelength UV radiation available on early Earth would affect overall reaction rates.

\section{Conclusions \label{sec:conc}}

Constraining the abundance of trace chemical species on early Earth is relevant to understanding the plausibility and guiding the development of proposed prebiotic chemistries. In this work, we have used box-model kinetic calculations to constrain the plausible range of \ce{NO_2^-}~and \ce{NO_3^-}~concentrations in oceans and ponds on prebiotic Earth. 

Prebiotic oceanic \ce{NO_{X}^{-}}~was likely much lower than calculated in previous work \citep{Wong2017,Laneuville2018} due to UV photolysis and reactions with \ce{Fe^{2+}}. Oceanic \ce{NO_{X}^{-}}~could only have built up to $\geq 1~\mu$M in an extremal realm of parameter space, in particular if the \ce{NO_X^-} supply flux was much higher than currently favored in the literature. Consequently, origins-of-life scenarios which require elevated \ce{NO_{X}^{-}}~in the ocean must invoke either an extremal planetary conditions, or specialized local conditions which concentrate \ce{NO_{X}^{-}}~above the oceanic mean. \ce{NO_{X}^{-}}~was not an inevitable part of the prebiotic milieu, and the prebiotic plausibility of oceanic origin-of-life scenarios can be improved by utilizing alternative feedstocks, e.g. an alternative electron donor for protometabolism \citep{Ducluzeau2009}.

Prebiotic \ce{NO_{X}^{-}}~could have build above oceanic levels in shallow ponds with large, fast-draining catchment areas. Such environments should have been extant but likely uncommon. In these environments, \ce{NO_{X}^{-}}~could have built up to prebiotically relevant levels ($\geq 1\mu$M) over a much broader range of planetary parameters than in the ocean, and in particular over most (but not all) of the proposed range of \ce{NO_X^-} supply flux. Consequently, \ce{NO_{X}^{-}}-dependent prebiotic chemistries which can function in shallow ponds (e.g., \citet{Mariani2018}) are prebiotically plausible, with the caveat that they impose requirements on the environment. Near-millimolar \ce{NO_{X}^{-}}~concentrations are possible if the \ce{NO_X^-} supply flux was at the upper end of what has been proposed in the literature ($\phi_{NO_{X}^{-}}\geq 6.5\times10^8 \text{cm}^{-2} \text{s}^{-1}$), and if the pond were cool and mildly acidic. This finding is in line with past work which suggests shallow lakes/ponds to be especially compelling venues for origin-of-life chemistry \citep{Mulkidjanian2012, Patel2015, Deamer2017, Ranjan2018}. We emphasize that our estimates are upper bounds; if a significant fraction of input NO$_X^-$ failed to go to NO$_X^-$, or reactions with other reductants present on early Earth were significant compared to the processes considered here, [NO$_X^-$ ] would have been proportionately lower.

For both oceanic and pond environments, \ce{NO_{X}^{-}}-dependent prebiotic chemistries that can function at lower [\ce{NO_{X}^{-}}] are more prebiotically plausible. Similarly, prebiotic chemistries that utilize \ce{NO_3^-}~are more plausible than those which utilize \ce{NO_2^-}, since most \ce{NO_{X}^{-}}~should be present as \ce{NO_3^-}~due to its greater stability.

Our analysis could be most improved by better characterization of \ce{NO_{X}^{-}}~reaction kinetics under prebiotically-relevant conditions, especially its reduction by \ce{Fe^{2+}}~and Mn$^{2+}$ and its reduction by UV in prebiotic natural waters (e.g., the extension of studies like \citet{Stanton2018} to NO$_X^-$). Studies with higher-dimensional models could also help determine if there exist areas in the ocean which should have concentrated \ce{NO_{X}^{-}}~above the oceanic mean, perhaps to prebiotically-relevant levels.

\acknowledgments
We thank Jonathan Toner for help with PHREEQC and insights regarding activity coefficients and diffusion. We thank Joshua Krissansen-Totton for a tutorial on his code, and discussions about thermal equilibrium. We thank Noah Planavsky, Clark Johnson, Scott Wankel, Taylor Perron, Sam Goldberg, William Bains, Craig Walton, and Sara Seager for helpful discussions. We thank Peter Gao, Matthie Laneuville, and Oliver Zafiriou for answers to questions. We thank the MIT Libraries for heroic efforts in uncovering very old, very obscure papers. We thank our referees and editor for feedback which substantively improved this paper. Z. R. T. and D. D. S thank the Harvard Origins of Life Initiative. This research has made use of NASA's Astrophysics Data System. This work was supported by grants from the Simons Foundation (SCOL  grant \# 495062 to S.R.; grant \# 290360 to D.D.S.), and support from an MIT startup (A.R.B.). P.B.R. thanks the Simons Foundation and Kavli Foundation for funding. 

The software used to carry out our calculations is available for validation and extension at \url{https://github.com/sukritranjan/nox.git}.

The authors declare no conflicts of interest with respect to the results of this paper.


%
\bibliography{nox.bib}
%




\end{document}